\documentstyle[pra,aps]{revtex}
\begin{document}
\draft
%
%
\title{Macroscopic Quantum Damping in SQUID Rings}
\author{L. Viola${}^{1,2}$\thanks{Vlorenza@mit.edu}, R. 
Onofrio${}^1$\thanks{Onofrio@padova.infn.it} and  T. 
Calarco${}^3$\thanks{Calarco@ferrara.infn.it} }
\address{${}^1$ Dipartimento di Fisica ``G. Galilei'', Universit\`a di Padova, 
and INFN, Sezione di Padova, \\ 
Via Marzolo 8, Padova, Italy 35131 \\
${}^2$ Department of Physics, Massachusetts Institute of Technology, \\
12-127, 77 Massachusetts Ave., Cambridge, MA 02139 \\ 
${}^3$ Dipartimento di Fisica, Universit\`a di Ferrara, 
and INFN, Sezione di Ferrara, \\ Via Paradiso 12, Ferrara, Italy 44100 } 
\date{\today}
\maketitle
%
%
\begin{abstract}
The measurement process is introduced in the dynamics of Josephson 
devices exhibiting quantum behaviour in a macroscopic degree of freedom. 
The measurement is shown to give rise to a dynamical damping 
mechanism whose experimental observability could be relevant to understand 
decoherence in macroscopic quantum systems.
\end{abstract}
%
%
\begin{center}
(Preprint {\sc dfpd/gp}/09, submitted to {\sl Physics Letters A})
\end{center}
%
\section{Introduction}
Since Schr\"odinger, both theoretical and experimental efforts have been 
devoted to understand whether quantum mechanical laws still work at the 
macroscopic level \cite{SCHRO}. 
In particular, many avenues are currently investigated to create coherent 
superpositions of macroscopically distinguishable states starting either
from microscopic degrees of freedom, as in the case of photons 
\cite{YURKE,PHOENIX,HAROCHE}, electrons \cite{MARTON}, neutrons 
\cite{MAIER}, atoms \cite{MLYNEK,MONROE} and molecules \cite{JANS}, 
or macroscopic ones, for instance by using superconducting circuits 
\cite{LEGGETT}. 
The common goal of these attempts is to grasp how the coherence  
properties, the crucial role of which is well established in the microworld, 
are washed out in the transition to the macroscopic realm, through what is 
commonly termed {\sl decoherence} \cite{ZEH,ZUREK}.
This last concept, invoked to explain why superpositions of macroscopic 
states are so fragile to make their actual observation far from being a 
trivial task, is present whenever the system under consideration is coupled to
the outside world, turning its evolution into an irreversible one. 
Among the possible sources of decoherence, the one attributable to the act of 
the measurement deserves special attention since, being any physical inquiry 
on the system performed via the coupling to an external probe, it is in 
principle unavoidable. In previous papers the measured induced decoherence 
has been analyzed in the case of {\sl microscopic} systems, namely atoms 
confined either in optogravitational cavities \cite{ONVIO} or in 
electromagnetic traps \cite{JC}. 
Observable quantities turn out to be modified 
giving rise to peculiar transient phenomena. In this work we extend the 
analysis to the dynamics of {\sl macroscopic} quantum systems 
such as the Josephson devices which, as emphasized by Leggett \cite{LEGGETT}, 
are promising candidates to study the coherence-decoherence transition 
by virtue of both their low operating temperature and the small intrinsic 
dissipations.  
In Section II we introduce the framework of measurement quantum 
mechanics and apply it to the measurement of energy in a radiofrequency 
superconducting quantum interference device (rf-SQUID).      
Decoherence signatures arising from the measurement process in the dynamics 
of the average magnetic flux are discussed in Section III. Final remarks on 
the feasibility and the differences from the microscopic systems complete the 
paper.
    
\section{Measurement quantum mechanics of a SQUID ring}

The system that will be considered hereafter is a superconducting closed 
loop interrupted by a weak link of the Josephson type, in short an rf-SQUID. 
Neglecting dissipation mechanisms, its quantum mechanical description 
can be modeled through the effective Hamiltonian \cite{BARONE}
\begin{equation}
\hat{H}= {\hat{P}^2_{\Phi} \over 2C} + \hat{V}(\Phi) = 
- {\hbar^2 \over 2C} {d^2 \over d \Phi^2} + 
{(\Phi-\Phi^{ext})^2 \over 2L} - {I_c \Phi_0 \over 2 \pi} \cos 
\bigg( 2\pi {\Phi \over \Phi_0 } \bigg) \;, 
\label{HAM}
\end{equation}
where $C$ is the weak link capacitance, $L$ the ring inductance, $I_c$ 
the critical current of the junction, $\Phi^{ext}$ is the external 
magnetic flux threading the ring and the flux quantum $\Phi_0=\hbar/2e 
\approx 2.07\cdot 10^{-15}$ Wb. In the circuit description (\ref{HAM}),  
the total magnetic flux $\hat{\Phi}$ and the total displacement flux 
(with units of electric charge) threading across the weak link capacitor, 
$\hat{P}_{\Phi}=-i\hbar d/d \hat{\Phi}$, play the role of generalized 
coordinate and momentum respectively. Notice that $C$ acts as the effective 
mass of the whole ring, treated as a single macroscopic quantum object 
evolving in $\Phi$-space under the potential $\hat{V}(\Phi)$. If the 
adimensional variable $x= (\Phi- \Phi^{ext})/\Phi_0 $ is introduced and a 
fixed value $\Phi^{ext}$ of the external flux is chosen fulfilling the 
relation $\Phi^{ext}/\Phi_0=(n+1/2)$, with $n$ integer to ensure 
$\Phi$-symmetry of the effective potential around $\Phi=\Phi^{ext}$, the 
latter can be rewritten as
\begin{equation}
V(x)= {\Phi^2_0 \over 2L} x^2 + {I_c \Phi_0 \over 2\pi} \cos (2\pi x)
\;. \label{POT}
\end{equation}
According to (\ref{POT}), monostable or multistable dynamics arises  
depending on the relative amplitude of the parabolic and periodic terms. 
In particular, a bistable regime is obtained if $1 < \beta < 5 \pi/2$, with 
$\beta= 2\pi \, L I_c/\Phi_0$. Under such conditions, the double well 
potential (\ref{POT}) can be approximated by a polinomial expansion, 
\begin{equation}
V(x)\simeq V_0 - {\mu \over 2} x^2 + {\lambda \over 4} x^4 \;, 
\hspace{2cm}\mu,\lambda >0\;, 
\label{WELL}
\end{equation}
being the parameters $\mu, \lambda, V_0$ related to the physical ones 
via the relationships:
\begin{eqnarray}
 & & \left\{ \begin{array}{ll}
             \mu & = 2\pi I_c \Phi_0 - \Phi_0^2/L\;, \\
         \lambda & = 4 \pi^3 I_c \Phi_0/3\;, \\
             V_0 & = I_c \Phi_0/2\pi = 3 \lambda/ 8\pi^4 \;. 
\end{array} \right.
\label{PARAM}
\end{eqnarray}    
The quartic potential approximation (\ref{WELL}), possessing two local 
minima at $\hat{x}_{\pm}= \pm \sqrt{\mu/\lambda}$ separated by a barrier 
of height $\Delta U = \mu^2/4\lambda$, has been widely exploited to model 
the macroscopic quantum tunneling of the flux through the junction 
\cite{LEGGETT}. One can easily check by inverting (\ref{PARAM}) that only the 
range of values $\mu < 3 \lambda/2 \pi^2$ is allowed in order the bistable 
potential (\ref{WELL}) corresponds to a physical SQUID. 
 
Let us now include the effect of a measurement process in the description of 
the SQUID ring. We will focus the attention on a situation where the 
measurement most clearly possesses a dynamical content, namely the continuous 
monitoring of a generic system observable $\hat{A}$. Significant progresses in 
modeling such a situation have been made in recent years through different 
approaches, all of which basically recognize that a measured system is not 
isolated but in interaction with a measurement apparatus. A detailed account 
of these approaches, together with the unified picture in which all of them 
can be incorporated, has been given elsewhere \cite{MQM}. Few remarks are in 
order to apply this general scheme to our particular problem at hand. Due to 
the fact that SQUIDs are macroscopic condensed matter objects and therefore 
two indiscernible rings (let alone an ensemble) cannot be constructed, 
measurement procedures are constrained to deal with just one SQUID and 
physical information emerges from averaging individual histories with the 
system each time prepared in the same initial macroscopic state. In the 
language of quantum measurement theory, such a procedure translates in the 
so-called {\sl selective a priori} measurement scheme, whereby the 
instantaneous result  of the measured observable corresponds to a particular, 
though {\it a priori} unknown, selection of the possible outcomes. Accordingly,
the deterministic Schr\"odinger evolution for the state vector is replaced by 
a stochastic one, where single realizations closely mimic 
the fluctuating output of each experimental run. Two equivalent approaches may
be used for unravelling the master equation, either a Quantum State Diffusion
picture where stochasticity acts continuously \cite{GISIN}, or 
the so-called Quantum Jumps approach resulting when stochasticity is chosen to
act occasionally \cite{CARMICHAEL}. 
{\sl Nonselective} measurement results are recovered by averaging the 
selective outcomes according to their probability distribution. On the other 
hand, a different description of the dynamics in the presence of the 
measurement, although not viable in practice when no ensemble is at disposal, 
has been proven to be equivalent to the previous one insofar as averaged 
(hence nonselective) predictions are required. Being in this case technically 
much more simpler, it will be adopted for the SQUID problem.    

The starting point for a direct description of nonselective measurements is 
the master equation, derived in the framework of open quantum systems, 
which rules the evolution of the density operator $\hat{\rho}(t)$ 
of the system under observation \cite{LINDBLAD},  
\begin{equation}
{d \hat{\rho}(t) \over dt}= - {i \over \hbar} [\hat{H},\hat{\rho}(t)] - 
{\kappa \over 2} [\hat{A},[\hat{A},\hat{\rho}(t)]]\:,   
\label{MASTER}
\end{equation}
where $\hat{H}$, given in (\ref{HAM}), describes the dynamics of the 
unmeasured closed system ($\kappa=0$) and the measured observable $\hat{A}$ 
is a so-called Lindblad operator representing the influence of the environment 
on the system. The parameter $\kappa$ expresses the strenght of the coupling 
between the measured system and the meter, that will be assumed to be 
time-independent throughout the derivation \cite{MQM}. Let us restrict the 
discussion to the case $\hat{A}=\hat{H}$, corresponding to a continuous 
measurement of energy, with $\kappa = \kappa_E$. Other situations in which 
the measurement process has been taken into account for a bistable potential 
have been previously discussed in \cite{SCHENZLE} and \cite{CALARCO} for 
continuous and impulsive monitoring respectively, but choosing as measured 
observable the (possibly generalized) coordinate itself, {\it i.e.} 
$\hat{A}=\hat{\Phi}$ in (\ref{MASTER}). Our choice, which reflects a viewpoint 
already discussed in \cite{ONVIO}, is instead guided by the demand to describe 
the effect of a measurement of one observable on the dynamics of another one 
which does not commute with the former. This also corresponds to the original 
philosophy of the quantum Zeno effect as introduced by Misra and Sudarshan, 
whereby measurements of the particle track were expected to affect the 
lifetime \cite{MISRA}. 
Moreover, the energy observable plays a peculiar role for all the models in 
which a fundamental diagonalization in the energy eigenstates (the so-called
{\sl intrinsic} decoherence) is predicted to occur and for which a formally
identical description applies \cite{MILBURN,PERCIVAL}.
The nonselective master equation (\ref{MASTER}) can be 
conveniently solved by introducing the representation of the measured 
observable eigenstates, $\hat{H}|n\rangle=E_n |n\rangle$, with 
$n=0,1,\ldots$ labelling the quantized energy levels of the Hamiltonian 
(\ref{HAM}) and density matrix elements 
$\rho_{nm}(t)= \langle n| \hat{\rho}(t) | m \rangle$. Eq. (\ref{MASTER}) is 
thus rewritten as
\begin{equation}
\dot{\rho}_{nm}(t)=-{i\over \hbar} (E_n-E_m)\rho_{nm}(t) - 
{\kappa_{E} \over 2} (E_n-E_m)^2 \rho_{nm}(t) \;, \label{DOT}
\end{equation}
whose general solution is 
\begin{equation}
\rho_{nm}(t)=\exp\left\{{-{i \over \hbar} (E_n-E_m)t -
{\kappa_E \over 2} (E_n-E_m)^2 t} \right\} \rho_{nm}(0)  \;.
\label{SOLUTION}
\end{equation}
This allows us to write down the exact density operator evolution for 
the system undergoing the measurement process in the form 
\begin{equation}
\hat{\rho}(t) = \sum_{nm} \rho_{nm}(0) \exp\left\{{-{i \over \hbar} 
(E_n-E_m)t - {\kappa_E \over 2} (E_n-E_m)^2 t} \right\}
|n \rangle \langle m| \;, 
\label{RHO}
\end{equation}
where the quantum decoherence effect accompanying the measurement is 
manifested as the vanishing of the off-diagonal density matrix elements at 
a rate proportional to the associated energy spacing \cite{ZUREK}. 
According to Eq. (\ref{RHO}), the evolution of the system has been 
altered with respect to the closed one, $\kappa_E=0$; it is important 
to understand how such modifications can manifest in observable 
properties of the system. Our approach, based on
a direct observation of the monitored quantum system with energy meter 
variables disregarded, can be considered as complementary to the adiabatic 
monitoring scheme already proposed in \cite{SPILLER}, where the quantum 
variables are instead traced out to leave a modified reduced evolution
of the apparatus. 

\section{Quantum damping of the average magnetic flux}

As a SQUID observable showing nontrivial dynamical behavior, let us consider 
the total average magnetic flux threading across the ring as a function of 
time,  
$\langle \hat{\Phi}(t)\rangle= 
\mbox{Tr}(\hat{\Phi}\hat{\rho}(t)) = \sum_{nm} \rho_{nm}(t) 
\langle m|\hat{\Phi}|n \rangle$. Owing to Eq. (\ref{RHO}), we get 
\begin{equation}
\langle \Phi(t) \rangle=\sum_{nm} \rho_{nm}(0)
\exp\left\{{-{i\over \hbar} (E_n-E_m)t - 
{\kappa_E \over 2}(E_n-E_m)^2t}\right\} \langle m |\hat{\Phi}|n \rangle \;.
\label{FLUX}
\end{equation}
At variance with the purely oscillatory behavior occurring when no 
measurement is performed, an exponential damping with state-dependent time 
constants $\tau_{nm}=2/[\kappa_{E}(E_n-E_m)^2]$ arises, for which no 
classical counterpart can be envisaged. After a transient whose duration 
depends, for a fixed $\kappa_E$, on the Bohr frequencies 
$\omega_{nm}=(E_n-E_m)/\hbar$ picked up by the initial configuration, the 
average flux dynamics becomes completely inhibited leading, irrespective of 
the initial state, to an average localization around the asymptotic value 
\begin{equation}
\lim_{t \rightarrow +\infty} \langle \hat{\Phi}(t) \rangle = \sum_n 
\rho_{nn}(0) \langle n|\hat{\Phi}|n \rangle =0 \;, \hspace{1cm} 
\kappa_E > 0\;,
\label{DAMPING}
\end{equation}
where odd parity of the flux operator has been exploited. 

In order to visualize the phenomenological features arising in the 
presence of quantum damping, we specialize the general description of 
Eq. (\ref{RHO}) by choosing some representative initial SQUID configurations 
of the system. A picture of the double well potential (\ref{WELL}) and its 
lowest energy eigenfunctions is reported in Fig. 1. Firstly, let the SQUID 
be initially prepared in a combination of the two lowest states, 
$|\Psi, t=0\rangle$ = ${1/\sqrt{2}} \,[ |0\rangle +|1\rangle ] = |L \rangle $, 
where $|0\rangle$ and $|1\rangle$ are the symmetric ground state and the 
antisymmetric first excited state respectively and the notation 
$|L\rangle$ has been introduced to stress that the dominant 
concentration of the state is in the left well. We will also consider the 
antisymmetric right-localized superposition state $|R\rangle$, defined by
${1 /\sqrt{2}}\, [ |0\rangle - |1\rangle ]$ = $|R \rangle $. 
Due to their localized nature, the states $|L\rangle$, $|R\rangle$ can 
be differentiated on a macroscopic level by the net left or right origin 
of the magnetic flux (also implying an opposite circulation sense for 
the superconducting current) and represent therefore a simple 
realization of two macroscopically distinguishable states. By evaluating 
the initial density matrix $\hat{\rho}(0)= |L\rangle \langle L|$, a damped 
oscillation is found from Eq. (\ref{FLUX}) for the average flux motion,
\begin{equation}
\langle \hat{\Phi}(t) \rangle = \exp\bigg( - {\kappa_E \over 2}  
(\hbar \omega)^2 t \bigg) \Re\mbox{e}\big\{ \langle 0 | \hat{\Phi} | 1 
\rangle \mbox{e}^{-i \omega t} \big\} \;, 
\label{DAMPED}
\end{equation}
where parity conservation has been again taken into account and $\omega_{10}
\equiv \omega =(E_1-E_0)/\hbar$ denotes the angular tunneling frequency, 
related to the height barrier and the zero-point energy $\hbar \omega_0$ in 
each well via a WKB estimate \cite{BARONE}:
\begin{equation}
\omega = \omega_0 \sqrt{{\Delta U \over \hbar \omega_0}} 
\exp\bigg\{ - {\Delta U \over \hbar \omega_0} \bigg\} \;. 
\label{TUNNEL}
\end{equation}
Transient phenomena in the average flux dynamics are ruled in this case by a 
time scale of the order $2/[\kappa_E (\hbar\omega)^2]$. 
Some interesting insights on the action 
of the measurement process can also be gained by depicting the evolution 
of the $|L\rangle$-state directly in the $LR$-representation, which is 
the natural one for tunneling analysis. This may be accomplished by 
rewriting the relevant projectors in (\ref{RHO}) according to the inverse 
transformation, {\it e.g.} 
$|0 \rangle\langle 0|= [ |L \rangle \langle L| + |R \rangle \langle R| +
|L \rangle \langle R| + |R \rangle \langle L| ] /2$, 
$|0 \rangle\langle 1|= [ |L \rangle \langle L| - |R \rangle \langle R| -
|L \rangle \langle R| + |R \rangle \langle L| ] /2$ and so on. As a 
result, the density operator of the measured system in the  
macroscopic $LR$-representation can be written as follows:
\begin{equation}
\hat{\rho}(t) \doteq \left( \begin{array}{cc}
1+\mbox{e}^{- \kappa_E (\hbar\omega)^2 t /2 } \cos \omega t  &
-i \mbox{e}^{- \kappa_E (\hbar\omega)^2 t /2 } \sin \omega t \\
 i \mbox{e}^{- \kappa_E (\hbar\omega)^2 t /2 } \sin \omega t &
1-\mbox{e}^{- \kappa_E (\hbar\omega)^2 t /2 } \cos \omega t  
                                      \end{array} \right). 
\label{RHOMACR}
\end{equation}
The off-diagonal density matrix elements make evident the quantum interference 
between the macroscopically distinguishable states $|L\rangle$ and 
$|R\rangle$, implying the existence of a Schr\"odinger cat state \cite{SCHRO}. 
According to (\ref{RHOMACR}), the decay of this macroscopic coherence 
represents the off-diagonal effect of the coupling to the meter. In addition, 
the measurement influence extends to the diagonal matrix elements, expressing 
the probability for left and right localization of the state respectively. 
Asymptotically, the tunneling mechanism becomes frozen, a behavior showing 
close similarities with the result quoted in \cite{KILIN}, where a dynamical 
suppression of tunneling in a symmetric double well potential arises as a 
consequence of an external perturbation - a coherent laser field driving 
transitions to an electronically excited bistable state. Of course, the same 
evolution given in (\ref{DAMPED}) has to be recovered using the density 
representation (\ref{RHOMACR}), 
\begin{eqnarray} 
\langle \hat{\Phi}(t) \rangle & = & 
\frac{\langle L| \hat{\Phi} |L \rangle}{2} \Big( 
1+\mbox{e}^{- \kappa_E (\hbar\omega)^2 t /2 } \cos \omega t \Big) +
\frac{\langle R| \hat{\Phi} |R \rangle}{2} \Big( 
1-\mbox{e}^{- \kappa_E (\hbar\omega)^2 t /2 } \cos \omega t \Big) 
\nonumber \\
 & = & \langle 0 |\hat{\Phi} |0 \rangle + 
\langle 0 |\hat{\Phi} |1 \rangle 
\mbox{e}^{- \kappa_E (\hbar\omega)^2 t /2 } \cos \omega t  \;. 
\label{DAMPED2}
\end{eqnarray} 

A more realistic preparation of initially localized states usually takes 
place by considering Gaussian flux wavepackets of the form 
\begin{equation}
\psi_0(\Phi)= {1 \over (\pi \sigma^2_\Phi)^{1/4} } 
\exp \bigg( - { (\Phi -\Phi_m)^2 \over 2 \sigma^2_{\Phi} } \bigg) = 
{1 \over \sqrt{\Phi_0} } {1 \over (\pi \sigma^2_x)^{1/4} } 
\exp \bigg( - { (x - x_m)^2 \over 2 \sigma^2_x } \bigg) \;, 
\label{GAUSSIAN}
\end{equation}
with the parameters $\Phi_m=x_m\Phi_0+\Phi^{ext}$ and $\sigma_{\Phi}=
\Phi_0 \sigma_x$ properly chosen to attain both the desired flux 
localization and a suitable initial average energy 
($\langle \psi_0 | \hat{H} |\psi_0 \rangle < \Delta U$ in the tunneling 
regime). Starting from 
\begin{equation}
\rho_{nm}(0)= \int \, d\Phi d\Phi' \varphi^\ast_n(\Phi)\psi_0(\Phi) 
\psi^\ast_0(\Phi') \varphi_m(\Phi') \;,
\label{INI}
\end{equation} 
the nonselective evolution of the average flux results from Eq. (\ref{FLUX}). 
By referring to \cite{ONVIO} for general considerations on the Gaussian case 
still applying here, a quantitative description is made available by resorting 
to numerical calculations. The required energy eigenfunctions and eigenvalues 
have been computed through a selective relaxation algorithm extensively 
described in \cite{PRETA}. A representative evolution arising from a Gaussian 
state maximally contributed by the first four energy eigenstates is depicted 
in the sequence of Fig. 2, where the effect of an increasing coupling to the 
meter proceeds from 2(a) to 2(f). A more resolved picture for the first three
cases follows in Fig. 3. The presence of the second doublet of almost 
degenerate eigenstates has been obtained by centering the Gaussian state 
around a value of magnetic flux intermediate between the ones corresponding to 
the left maxima of the eigenfunctions (Fig. 1). The transient strongly depends 
on the initial state and exhibits, as already remarked, many decay timescales.
When $\kappa_E > 0$, after a time $\tau$ only the oscillations involving 
levels with energies $|E_n-E_m| <\sqrt{2/(\kappa_E\tau)}$ will survive. 
More in general, critical values of the measurement coupling constant can be
introduced for each oscillation pattern as\footnote{For analogous definitions
of critical measurement couplings in atomic systems see for instance 
\cite{JC,MQM}.}
\begin{equation}
\kappa^{crit}_{nm}= {1 \over {T_{nm} (E_n-E_m)^2 } }= 
{1 \over {h (E_n-E_m) } } \;, 
\label{CRITICAL}
\end{equation}
where $T_{nm}=2\pi/\omega_{nm}$ are the periods of the Bohr oscillations. Each 
critical value rules the $nm$-subdynamics of the SQUID magnetic flux, giving 
rise to deviations from the oscillatory closed system evolution for 
$\kappa_E > \kappa^{crit}_{nm}$. Since the smaller energy splitting is due to 
the ground state ($|E_1 - E_0| > |E_n-E_m|$ $ \forall n,m$), the overall 
dynamics will be affected if $\kappa_E > \kappa^{crit}_{10}$. By varying 
$\kappa_E$ it is thus possible to observe qualitatively different initial 
transients, selecting for instance the oscillations at the frequencies 
$\omega_{32}$ and $\omega_{10}$ (Figs.~2(b) and 3(b)), or the latter only 
(Fig.~2(c--f)). Further increase in $\kappa_E$ causes the oscillations
to disappear at all, Fig.~2(c-f). 

Few remarks are worthwhile. Similar considerations can be in principle 
repeated for the evolution of any ensemble average quantity, the explicit 
solution of the appropriate stochastic Schr\"odinger equation (or Quantum 
Jumps equation) being required to calculate individual SQUID behavior. 
Note, however, that the transient nature of the expected phenomena 
would not allow us to exploit time averages of the monitored quantities 
sampled over a single stochastic trajectory, the former being 
only viable if the observed system is at thermal equilibrium as discussed in 
\cite{SUSSEX}. Indeed, having thermal fluctuations been neglected in our case, 
the same argument given in \cite{GISIN} can be applied to show that, for a 
SQUID density matrix evolving according to Eq. (\ref{RHO}), convergence of 
individual runs to energy eigenstates is expected.   
Conceptually, the damping mechanism operating in the nonselective ensemble 
evolution of average properties can then be also interpreted as 
due to the random dephasing between different histories associated to the 
selective evolutions of the same properties - both dephasing and damping being 
signatures of irreversible dynamics. In addition, in order to make the 
predicted behavior qualitatively similar to the one occurring in practice, the 
physical parameters have to be chosen in such a way that both bistable and 
quantum tunneling regimes are guaranteed by the potential energy (\ref{POT}). 
According to the values of $\mu, \lambda$ corresponding to the Fig. 1, an 
ultra small capacitance SQUID ring as used in \cite{PRANCE} has been modeled, 
$C \sim 10^{-16}$ F, leading to $L \sim 10^{-11}$ H, $I_c \sim 20$ $\mu$A. For 
such a configuration, which is not too distant from routine design, thermal 
activation can be made negligible at relatively high operating temperatures, 
{\it e.g.} $k_B T/ \Delta U \sim 6 \cdot 10^{-3}$ at $T=4$ K.

\section{Discussion and conclusions}

A quantum damping due to the interaction between a macroscopic quantum system 
and an informational environment has been discussed in the case of a SQUID 
ring undergoing continuous nonselective measurements of energy. This damping 
has no classical analog and is strongly imprinted by the decoherence process 
originating in the act of measurement. 

The predicted effect can be made observable for instance by measuring the 
dynamics of the average magnetic flux in the SQUID or any other observable 
which does not commute with the Hamiltonian (\ref{HAM}). Since the system 
under observation is a single macroscopic object the procedure must consist 
in repeating the cycle of measurements, each characterized by the following 
steps. Firstly, the system is prepared in a flux state, say a Gaussian, 
centered around one of the two minima for instance by properly adjusting the 
external magnetic flux. At the same time the energy meter is turned on. The 
total magnetic flux is measured after a given duration $t$ by means of an 
instantaneous von Neumann measurement. The cycle is repeated again at constant 
$t$ allowing to build the magnetic flux distribution and therefore the average 
magnetic flux. The whole procedure is repeated again by sweeping the time $t$,
allowing a comparison with the predicted behaviour (\ref{FLUX}). Concerning 
continuous measurements of energy, to our knowledge no experimental technique 
is by now available. The closest achievement has been reported by the Sussex 
group regarding the spectroscopy of quantum mechanical SQUID rings 
\cite{PRANCE}. The kind of measurement required here is however quite 
different, since a  dynamical monitoring of energy is demanded. In other words,
even if the knowledge of the energy levels is a prerequisite, a monitoring of 
the populations in each energy eigenstate is also necessary. For states formed 
by the lowest two levels, this could be achieved by coupling the two-level 
dynamics to a third level in a way similar to the one exploited to monitor the 
electronic populations in the study of atomic quantum Zeno effect \cite{ITANO}.
In the SQUID problem, the lack of available schemes does not allow to make 
quantitative comparison between the predicted time decay due to quantum 
damping and the decays times due to classical sources of decoherence 
associated to intrinsic dissipations \cite{BEN}. In addition, if fundamental 
energy localization is proven to occur in nature, the detection of 
measurement induced energy localization will only be possible if the 
measurement coupling $\kappa_E$ is sufficiently large compared to the minimum
step time associated to intrinsic decoherence \cite{MILBURN,PERCIVAL}.
In any case, we guess that its actual observability is obtainable for values 
of the parameters not far from those required to make intrinsically quantum 
phenomena, such as macroscopic quantum coherence \cite{COSMELLI} or temporal 
Bell inequalities \cite{LEGGETT,TESCHE}, detectable. 
It is also worth to point out that the predicted damping can be also 
implemented in a more quantitative way by considering tunneling phenomena at 
the microscopic level, for instance in the spectroscopy of molecular systems 
as considered in \cite{KILIN}.
On the other hand, in this case the possibility to test decoherence of a 
macroscopic degree of freedom is lost. 

The relevance of quantum damping in post-modern quantum mechanics is twofold. 
Firstly, it is an inequivocable test of the validity of current models on the 
quantum-to-classical transition which identify decoherence induced by the 
opening of the quantum system to the external world as the key concept. 
The possibility to prepare states in such a way to produce controllable decay 
patterns can support evidence for or against quantitative predictions of 
decoherence. Moreover, the decoherence issue in SQUID rings is crucial to 
exploit superconducting circuits as quantum computers \cite{BENNETT}.
The model presented here can be considered a highly idealized one describing 
the ultimate source of decoherence due to the informational input-output 
operations performable on superconducting quantum gates.

\acknowledgments 
It is a pleasure to thank Mark F. Bocko for a critical reading of the 
manuscript. This work has been partially supported by INFN, Sezioni di Ferrara 
and Padova, and INFM, Sezione di Roma 1. One of us (L.V.) thanks the 
Massachusetts Institute of Technology for the hospitality during the 
completion of this paper.

%
%
\begin{figure}
\caption{\label{fig1}
Plot of the double well potential $V(x)-V_0$ of Eq. (\protect\ref{WELL}) 
for the parameter choice $\mu=1.80487(1)$ eV, $\lambda =14.73360(1)$ eV, 
resulting in local minima at $|\hat{x}_{\pm}|=0.35$ and height barrier \
$\Delta U= 0.055274$ eV. The four lowest energy normalized eigenfunctions 
$\varphi_n(x)$ $(n=0,\ldots,3)$ are also shown together with their
energies $E_0=-0.0440591$ eV, $E_1= -0.0440585$ eV,
$E_2=-0.0231600$ eV, $E_3= -0.0230991$ eV. } 
\end{figure}
  
\begin{figure}
\caption{\label{fig2}
Predicted time dependence, in units of the tunneling period $T_{01}$, of the 
average magnetic flux for different 
values of the measurement coupling relative to the critical values  
$\kappa^{crit}_{10}$ and $\kappa^{crit}_{32}$ and for an initial Gaussian 
wavepacket having $x_m=-0.27$ and $\sigma_x=0.06$. The state is contributed 
up to $\approx 97\%$ by the two lowest energy doublets. 
(a) $\kappa_E=0;\,$
(b) $\kappa_E=10^{-2}\kappa^{crit}_{32} =10^{-4}\kappa^{crit}_{10};\,$
(c) $\kappa_E=10^{-1}\kappa^{crit}_{32} =10^{-3}\kappa^{crit}_{10};\,$
(d) $\kappa_E=10\kappa^{crit}_{32} =10^{-1}\kappa^{crit}_{10};\,$
(e) $\kappa_E=10^{2}\,\kappa^{crit}_{32} =\kappa^{crit}_{10};\,$
(f) $\kappa_E=10^{3}\,\kappa^{crit}_{32} =10\,\kappa^{crit}_{10}$.
The dark band in (a)--(c) around the tunneling motion with 
$\omega_{10}$ is due to the high frequency secondary oscillations 
picked up by the initial state. } 
\end{figure}

\begin{figure}
\caption{\label{fig3}
Same as Fig.~2 on a time scale expanded by an order of magnitude for the 
previous cases (a), (b) and (c). In (a) all the Bohr frequencies are present; 
in (b) only the slower oscillations at $\omega_{10}$ and $\omega_{32}$ survive,
with $\omega_{32}/\omega_{10} \sim 10^{-2}$. The $\omega_{32}$ component 
disappears within a tunneling period for a larger value of $\kappa_E$, as in 
(c). } 
\end{figure}

\end{document}